# Use of the modified Ginzburg-Landau equations in high temperature superconductors

**L.N Shehata(*) and H.M.Taha**
**Department of Mathematics and Theoretical Physics,**
**Atomic Energy Authority, Egypt,**
**AEA Post Office, Cairo, Egypt**

**Abstract**: The modified Ginzburg-Landau equations are used to study some fundamental problems for the high temperature superconductors. The derivations of these fundamental equations are summarized and the layered features of the sample were taken into consideration and the relation between the coherence length and the separation distance of the layers. The domain wall energy, the maximum supercurrent and the parallel critical field of high temperature superconducting thin film were evaluated. Comparison with previous works is given.

## 1 Introduction

In 1950, and on the basis of the Landau's theory of second – order phase transition, Ginzburg and Landau [1] (GL) successfully proposed their GL theory to describe the properties of the conventional ( or low temperature superconductors (LTS) ) near the transition temperature $T_c$. Later on, in 1957, the microscopic theory [2] of Bardeen – Cooper - Schriefer (BCS) was developed and the equivalence of the two theories as $T \rightarrow T_c$ was proven by Gor'kov [3].This shows that the suggested GL order parameter and the coefficients of the free energy functional can be calculated from the microscopic mean field theory. On the other hand, the GL theory has successfully overcome the London electrodynamics difficulties, e.g., the origin of the positive surface energy.

After the discovery of the first high temperature superconducting (HTS) compound La - (Ba, Sr)-Cu-O (at $T_c$ ~ 40 K) by Bednorz and Mueller [4] in 1986, several attempts were done to understand the real microscopic mechanism of the behavior of these novel materials. Nevertheless, the GL theory remains very useful and successfully to describe and explain the behavior of HTS regardless of the pairing mechanism.

On the basis of Landau's theory of second-order phase transition, Ginzburg and Landau were able to construct for conventional superconductors the free energy functional in terms of the powers of a complex order parameter $\psi(\vec{r})$. It is known that the region of thermal fluctuations of the order parameter near the mean-field theory critical temperature, $T_c$, for these LTS materials is very small [5] and could be neglected. Because of this, the mean-field theory was not able to explain the reason of other second-order phase transitions, e.g., superfluidity helium, liquid crystals, etc. Theoretically the range of temperature around $T_c$ and within which the fluctuations are effective, for pure SC, is $|T - T_c| < 10^{-14} T_c$, whereas, for alloys (type II SC) it is $|T - T_c| < 10^{-7} T_c$. These values are negligible and could be not achieved experimentally [6], and therefore, it was neglected in the original GL work for LTS.

(*) Corresponding author: e- mail: louisns@frcu.eun.eg

In the new HTS the coherence length $\xi(T)$ near $T_c$ extrapolated to T = 0, $\xi(0)$, is of the order of the interatomic distance d. Early, the theory of superfluidity was proposed for $^4He$, where $\xi(0) \sim$ d. In this theory, the GL free energy functional for the order parameter and with modified temperature dependence of the coefficients was taken as the effective Hamiltonian [7,8]. Several works, therefore, stressed on the possibility of using the same modified GL theory for the case of HTS [9-12]. The proposed form of the MGL effective free energy functional expansion in terms of the complex order parameter $\psi(\vec{r})$, with modified expansion coefficients, the term $|\psi|^6$ with temperature independent constant coefficient and with the effective mass tensor $m^*_{ik}$ to describe the anisotropy of HTS, was used as a modified form of the free energy functional for HTS.

The purpose of the present study is to use these modified Ginzburg - Landau (MGL) equations, obtained from the modified free energy, to solve relevant problems for HTS. Therefore, our work is organized as follows. In sec.2 a brief preview of the derivations of the MGL equations will be demonstrated. We will use the MGL equations [10] to solve three fundamental problems for HTS. The first one concerns the surface energy between normal and superconducting phases and whether the solutions of these equations in both cases of LTS and HTS are the same or not (sec. 3.1). In sec.3.2 we will consider the problem of thin film to calculate the maximum critical current that could be carried by this film or wires of HTS. The critical parallel field in the thin film will be given in sec. 3.3. In sec.4 some concluding remarks and conclusions will be given.

## 2 Brief preview on the derivations of the MGL equations

The general theory of Landau's second order phase transition is based on the existence of a complex order parameter $\psi(r)$, which tends to zero at the transition temperature and its temperature dependence is $\propto (T - T_c)/T_c$. The free energy of the system may be expanded in terms of the even powers of $|\psi|$ and the coefficients of the expansion are regular functions of the temperature. In its original form for conventional superconductors the free energy expansion ends at the forth power of $|\psi|$ and the temperature dependent of the $|\psi|^2$ coefficient, $\alpha(T)$, takes the form $\alpha(T) = \alpha_o \, \tau$, $\tau = (T - T_c)/T_c$. The coefficient of $|\psi|^4$ was taken to be temperature independent constant.

Experimental results [13] ensure the anisotropy structure of the HTS. In this case the scalar effective mass $m^*$ is replaced by a mass tensor $m^*_{ik}$. In the present study we will consider the case of uniaxial anisotropy, i.e., $m_x = m_y = m_{\parallel}$, where $m_{\parallel}$ is the parallel mass component to the sample surface, and $m_z = m_{\perp}$- the perpendicular mass component to the sample surface. The HTSs as we know are characterized by the layered structure. However, the relation between the normal component of the coherence length $\xi_{\perp}(T)$ and the layers separation, s, determines the relevance of the layered details to be considered in the system. As long as the coherence length component perpendicular to the layers $\xi_{\perp}(T)$ (or $\xi_z(T)$) is large compared with the distance between the layers, s, the layered details will be irrelevance and one can use the effective mass approximation [14-16] to write the following form for the MGL free energy functional, i.e.,

$$F\{\psi\} = \int_V F_s \, dV , \tag{1}$$

where, (*compare with ref.[21]*),

$$\begin{aligned} F_s = F_n &+ \frac{c_o}{2} T_c \tau^2 \ln|\tau| + \alpha(T)|\Psi(r)|^2 + \tfrac{1}{2}\beta(T)|\Psi(r)|^4 + \tfrac{1}{3} g_o |\Psi(r)|^6 + \frac{\boldsymbol{B}^2}{8\pi} \\ &+ \frac{\hbar^2}{2}\left[\frac{1}{m_{\parallel}}\left|\left(i\boldsymbol{\nabla}_{\parallel} + \frac{2e}{\hbar c}\boldsymbol{A}_{\parallel}\right)\Psi\right|^2 + \frac{1}{m_{\perp}}\left|\left(i\frac{\partial}{\partial z} + \frac{2e}{\hbar c}A_z\right)\Psi\right|^2\right] \end{aligned} \tag{2}$$

In the above formula, the first term $F_n$ in the right hand side is the free energy of the normal phase. The second term gives that part of energy, which should be added to the free energy due to the order parameter fluctuations (for more details see ref. [8, 10, and 17]). $\alpha(T) = \alpha_o \, \tau \, |\tau|^{1/3}$, and $\beta(T) = \beta_o \, |\tau|^{2/3}$ are the temperature dependent coefficients of the second and fourth power of the order parameter, respectively, and $g_o$ is a temperature independent coefficient of $|\psi|^6$. $\boldsymbol{B} = \mathrm{curl}\,\boldsymbol{A}$, where $\boldsymbol{B}$ is the microscopic magnetic induction inside the SC and $\boldsymbol{A}$ is its associated vector potential [10].

In the absence of the fields and the spatial variations of $\psi(r)$, $F_s$ will have minimum value, $F_{se}$, which is associated to the equilibrium value of the order parameter $\psi_e$, i.e.,

$$\psi_e^2(T) = (1 - M)\frac{|\alpha(T)|}{\beta(T)} , \tag{3}$$

and

$$M = \frac{g_o}{|\alpha(T)|}\psi_e^4(T) \tag{4}$$

Here, the dimensionless parameter M characterizes the existence of the term $|\psi(r)|^6$ in equation (2). This term was neglected in the case of LTS. From esq. (3) and (4) it is easy to verify that $F_s$ is a minimum when $\alpha(T)$ is negative and $\beta(T)$ is positive, where both of them are temperature dependent coefficients, while $g_o$ is a positive constant, and M satisfies the inequality $0 \le M < 1$. Therefore, for LTS, M = 0 and for HTS $M = 1/2$. Expression (3) may be used to find the condensation energy, which takes the form

$$F_n - F_s = \frac{H_c^2}{8\pi} = (1 - M)\left(1 + \frac{M}{3}\right)\frac{\alpha^2(T)}{2\beta(T)} \qquad (5)$$

To obtain the MGL equations the free energy $F_s$, eq. (2), is minimized with respect to $\psi^*$ and $\boldsymbol{A}$. Subsequently, we get the following two MGL equations, the solution of which will give the values of $\psi$ that correspond to the minimum of $F_s$ . The MGL equations reds,

$$\frac{\hbar^2}{2}\left[\frac{1}{m_\parallel}\left(i\boldsymbol{\nabla}_\parallel + \frac{2e}{\hbar c}\boldsymbol{A}_\parallel\right)^2\Psi + \frac{1}{m_\perp}\left(i\frac{\partial}{\partial z} + \frac{2e}{\hbar c}A_z\right)^2\Psi\right] + \alpha\Psi + \beta\Psi|\Psi|^2 + g_o\Psi|\Psi|^4 = 0 \qquad (6)$$

$$(\boldsymbol{curl\ B})_\parallel = \frac{4\pi\hbar e}{i m_\parallel c}\left(\psi^*\nabla_\parallel\psi - \psi\nabla_\parallel\psi^*\right) - \frac{16\pi e^2}{m_\parallel c^2}|\psi|^2\boldsymbol{A}_\parallel \qquad (7.1)$$

$$(\boldsymbol{curl\ B})_\perp = \frac{4\pi\hbar e}{i m_\perp c}\left(\psi^*\frac{\partial\psi}{\partial z} - \psi\frac{\partial\psi^*}{\partial z}\right) - \frac{16\pi e^2}{m_\perp c^2}|\psi|^2 A_z \qquad (7.2)$$

The Maxwell equation should be added to these formulas, i.e.,

***div B*** = 0 (7,3)

The boundary conditions at the surfaces of the superconductor for the MGL equations are

(1) $$\left(i\boldsymbol{\nabla}_\parallel + \frac{2e}{\hbar c}\boldsymbol{A}_\parallel\right)\cdot\hat{n} = 0 \qquad (8)$$

(2) $$\left(i\frac{\partial}{\partial z} + \frac{2e}{\hbar c}A_z\right).\hat{n} = 0 \qquad (9)$$

$\hat{n}$, here, is the normal component of the unit vector to these boundaries .The first MGL equation (6) differs from the first usual GL equations [17] by the modified temperature dependent coefficients $\alpha(T)$, $\beta(T)$ and the term $g_o|\psi|^4\psi$. These equations will be used in the following sections to explore their effectiveness to solve relevant problems for HTSs.

## 3 Use of the MGL equations

### 3.1 The Domain Wall Energy

The concept of the domain wall energy, or the surface energy, was firstly introduced by London [20]. In his concept the Meissner state cannot have the lowest energy without the existence of domain wall energy. Later, it was stated by Abrikosov [19] that the magnetic flux can penetrate a type II SC in the form of flux lines each carrying a quantum of magnetic flux. In electromagnetic London theory the vortex core was treated as a singularity. The phenomenological GL theory well describes many properties of the flux lines. It is now understood that the mixed state in HTS is an energetically favorable state because of the presence of the flux lines lattice, which subdivides the superconducting bulk into alternate normal and superconducting domains.

For simplicity we will consider a one-dimensional, infinite isotropic SC sample with a boundary plane z = 0 in an external magnetic field $H_0$ directed along the SC boundary. It will be assumed that $H_0 = H_c$, the value of the thermodynamic critical field. Taking into account the reduction of the magnetic energy due to the penetration of the field, the surface energy will then be the difference between the energy $F_p$ and the condensation energy, that would appear if the sample were entirely superconducting, where $F_p$ is given by

$$F_p = \int dv \Big[ \alpha(T) |\Psi(r)|^2 + \tfrac{1}{2}\beta(T) |\Psi(r)|^4 + \tfrac{1}{3} g_o |\Psi(r)|^6 + \frac{\boldsymbol{B}^2}{8\pi} - \frac{\boldsymbol{B} H_c}{4\pi} + \frac{H_c^2}{4\pi} + \frac{\hbar^2}{2m_\parallel}\left(\boldsymbol{\nabla}_\parallel - \frac{2ie}{\hbar c}\boldsymbol{A}_\parallel\right)^2 \Psi + \frac{\hbar^2}{2m_\perp}\left(\frac{\partial}{\partial z} - \frac{2ie}{\hbar c} A_z\right)^2 \Psi \Big] \tag{10}$$

In the above formula the induction $\boldsymbol{B} \equiv (B_x(z), 0, 0)$. Therefore, the domain wall energy $\sigma_{ns}$ per unit area will have the form,

$$\sigma_{ns} = \frac{F_p - \int \left( H_c^2 / 8\pi \right) dv}{s} \tag{11}$$

Where, s denotes the area of the SC and the second integral in the above definition, as was stated above gives the reduction in the free energy if the system were entirely SC. This reduction is described by the condensation energy $E_{cond}$, where,

$$E_{cond} = \int_{-\infty}^{+\infty} \frac{H_c^2}{8\pi} dV. \tag{12}$$

In order to calculate the wall energy $\sigma_{ns}$, for HTS sample, we have to use the MGL equations (6) and (7) and introduce the dimensionless quantities defined as [17]

$$\psi = \psi_e f, \quad r = \frac{r}{\lambda_\parallel}, \quad \boldsymbol{A} = \frac{2e}{c\hbar} \xi_\parallel(T) \boldsymbol{A}, \quad \boldsymbol{B} = \frac{2e}{c\hbar} \xi_\parallel(T) \lambda_\parallel(T) \boldsymbol{B} \tag{13}$$

The surface energy and the MGL equations will take the following forms in terms of the dimensionless notation by taking into account eq.(3),. The surface energy reads,

$$\sigma_{ns} = \frac{H_c^{\,2}\lambda_{\Box}}{4\pi\left(1+\frac{M}{3}\right)} \int_{-\infty}^{+\infty} dz\left[-f^2+\frac{1\text{-}M}{2}f^4+\frac{Mf^6}{3}+\frac{1}{\gamma\kappa^2}\frac{d}{dz}\left(f\frac{df}{dz}\right)- -\frac{1}{\gamma\kappa^2}f\left(\frac{d^2f}{dz^2}\right)+\frac{1}{f^2}\left(\frac{dB}{dz}\right)^2+\left(B_c\text{-}B\right)^2\right] \tag{14}$$

The MGL equations take the forms,

$$\frac{-\nabla_{\Box}^{2}}{\kappa^2}f_o - \frac{1}{\gamma\kappa^2}\frac{\partial^2 f_o}{\partial z^2} + \frac{1}{f_o^{\,3}}\left[\boldsymbol{curl\ B}\right]^2+\frac{\gamma-1}{f_o^{\,3}}\left[\left(\boldsymbol{curl\ B}\right)_z\right]^2 = =\left[1\text{-}\left(1\text{-}M\right)f_o^2\text{ - }M\,f_o^4\right]f_o \tag{15}$$

$$f_o^{\,2}\,\boldsymbol{B}\text{ - }f_o^{\,2}\frac{\gamma\text{-}1}{\gamma}\,\boldsymbol{curl}\left(\boldsymbol{0},A_{oz}\right)=\frac{2}{f_o}\ \text{grad}\,f_o\times\boldsymbol{curl\ B}-\ \boldsymbol{curl\ curl\ B} \tag{16}$$

Where,

$$f = f_o\ \exp\left(i\varphi\right) \tag{17}$$

$f_o$ is the modulus of the reduced order parameter and $\varphi(r)$ is its phase, and ***A*** satisfies the gauge invariance

$$\boldsymbol{A} = \boldsymbol{A_0} + \frac{\nabla\varphi}{\kappa} \tag{18}$$

where,

$$\boldsymbol{A_o} = \left(\boldsymbol{A_{o\Box}}, A_{oz}\right) \tag{19}$$

The thermodynamic critical field in reduced units $B_c$ reads

$$B_c = \frac{1}{\sqrt{2}}\left(1+\frac{M}{3}\right)^{1/2} \tag{20}$$

The boundary conditions necessary to solve eqs. (15) and (16) are:

In the normal region, $z \le 0;$ we require $f = 0$ and $B=B_c$, (21)

In the SC region, $z > 0;\ f = 1,\ \frac{df}{dz} = 0,$ and B = 0. (22)

By using these boundary conditions, and by taking into account the boundary behaviors of f and ***B***, eqs (15) and (16) may be rewritten in the form:

$$-\frac{1}{\gamma\kappa^2}\frac{\partial^2 f}{\partial z^2}+\frac{1}{f^3}\left(\frac{dB}{dz}\right)^2=\left[1-(1-M)\,f^2-M\,f^4\right]f \tag{23}$$

$$f^2B=-\frac{2}{f}\frac{df}{dz}\frac{dB}{dz}+\frac{d^2B}{dz^2} \tag{24}$$

$$Af^2=\frac{dB}{dz} \tag{25}$$

After tedious calculations, but not difficult we obtain the following results

$$\sigma_{ns}=2\lambda\frac{H_c^{\,2}}{8\pi}\delta\ , \tag{26}$$

where,

$$\delta=\frac{1}{\left(1+\frac{M}{3}\right)}\int_{-\infty}^{+\infty}dz\left[(B_c\text{-}B)^2-\frac{1}{2}(1\text{-}M)f^4-\frac{2}{3}Mf^6\right] \tag{27}$$

The type of the SC is determined according to the sign of $\delta$, which defines the character of the phase partition in the sample. A type I SC is characterized by a positive $\delta$, while a negative $\delta$ characterizes the type II SC.

In type I SC the magnetic field obeys Meissner state, in which ***B*** = 0 (up to $H_0 = H_c$), and therefore, eq.(23) reduces to,

$$-\frac{1}{\gamma\kappa^2}\frac{\partial^2 f}{\partial z^2}=f-(1-M)\,f^3-M\,f^5 \tag{28}$$

This equation should be solved together with the boundary conditions given by eqs. (21) and (22)); this gives

$$\frac{df}{dz}=+\sqrt{\gamma}\,\kappa\,\left(1-f^2\right)^{1/2}\left(B_c^{\,2}-\frac{1}{2}\left(1-\frac{M}{3}\right)f^2-\frac{1}{3}M\,f^4\right)^{1/2} \tag{29}$$

Since f = 0 and $B = B_c$ as z < 0, it is easy to verify that in this case $\sigma_{ns}=0$. On the other hand, in the region, where z > 0, B = 0, $f\neq 0$ and $\delta$ is evaluated numerically with the use of eq.(27) to give;

$$\sigma_{ns}=\frac{\zeta(T)}{\sqrt{\gamma}}\frac{H_c^{\,2}}{8\pi}\Delta_I \tag{30}$$

Where,

$$\Delta_I = \int_0^1 \left(1+f^2+\frac{4M}{3+M}f^4\right)\left(\frac{1}{2}+\frac{M}{3}\left(\frac{1}{2}+f^2\right)\right)^{-1/2} df \qquad (31)$$

Therefore, for type I SC, M = 0 (LTS), $\Delta_I = 1.887$ and in this case $\kappa = \lambda_{\square}/\xi_{\square} \;\square\; 1$.

For type II SC $\lambda_{\square}(T) \;\square\; \zeta_{\square}(T)$, i.e., $\kappa \;\square\; 1$ and therefore, the first term on the left hand side of the equation (23) may be neglected, and equations (24) and (25) will have the forms

$$\frac{dB}{dz} = -f^2\left[1-(1-M)f^2-Mf^4\right]^{1/2} \qquad (32)$$

$$B(z) = -\frac{d}{dz}\left[1-(1-M)f^2-Mf^4\right]^{1/2} \qquad (33)$$

The last equation gives

$$B = \left[\frac{f\left(1-M+2Mf^2\right)}{\left[1-(1-M)f^2-Mf^4\right]^{1/2}}\right]\frac{df}{dz} \qquad (34)$$

The magnetic induction B may be found from equations (32) and (34) by using the boundary conditions (21) and (22) i. e.,

$$B^2 = B_c^2 - \frac{1}{2}(1-M)f^4 - \frac{2}{3}Mf^6 \qquad (35)$$

The last equation shows that B = 0 as f = 1, and B = $B_c$ as f = 0, as it should be. For LTSs M = 0 and $B_c = 1/\sqrt{2}$ (see eq.(20)), and for HTSs (M = 1/2) $B_c = 0.7638 > 1/\sqrt{2}$. It is easy to show that for type II SCs

$$\sigma_{ns} = -\lambda_{\square}\frac{H_c^2}{8\pi}\Delta_{II} \qquad (36)$$

Where $\Delta_{II} \approx 2$ for M = ½ (HTS) and the wall energy in this case is always negative as it should be. The interface indicator $\delta$ is zero and $\kappa = \kappa_c$, where

$$\kappa_c = \sqrt{\frac{1}{2}(1-M)\left(1+\frac{M}{3}\right)}, \qquad (37)$$

which gives $\kappa_c = 0.54 < \frac{1}{\sqrt{2}}$ for M = ½ and $\kappa_c = \frac{1}{\sqrt{2}}$ for M = 0.

The critical value of GL-parameter $\kappa_c$ should be $\frac{1}{\sqrt{2}}$, which is obtained as M = 0. This value does not indicate the phase partition corresponds to LTS or HTS. However, in general, for $\kappa > \kappa_c = \frac{1}{\sqrt{2}}$ we have type II HTS and for $\kappa < \kappa_c = \frac{1}{\sqrt{2}}$ the SC belongs to the LTSs. Therefore, the values of M are $0 \le M \le \frac{1}{2}$.

### 3.2 Maximum supercurrent in HTS thin film

In this section, we consider an important simple example in which exact analytic solution of MGL equations is possible. An interesting example is that in which $|\psi|$ has the same value everywhere (without any substantial variation), e. g. a thin film or wire with thickness $d \ll \xi(T)$. Thus, put $\psi(r) = |\psi|\ e^{i\,\varphi(r)} = \psi_e\ f\ e^{i\,\varphi(r)}$, where, $f = |\psi|/\Psi_0$ . Therefore, one may write the free energy density in the following form

$$F = \int dv\Big[F_n + \alpha(T)|\Psi(r)|^2 + \tfrac{1}{2}\beta(T)|\Psi(r)|^4 + \tfrac{1}{3}g_o|\Psi(r)|^6 + \frac{1}{2m_\parallel}\boldsymbol{V}_\parallel^2|\Psi|^2 + \frac{1}{2m_\perp}V_\perp^2|\Psi|^2\Big], \tag{39}$$

and for the currents

$$\boldsymbol{J}_{\parallel s} = \frac{2e}{2m_\parallel}|\Psi|^2\left(\hbar\boldsymbol{\nabla}_\parallel\theta - \frac{2e}{c}\boldsymbol{A}_\parallel\right) = 2e|\Psi|^2\ \boldsymbol{V}_{\parallel s}, \tag{40}$$

$$J_{\perp s} = \frac{2e}{2m_\parallel}|\Psi|^2\left(\hbar\frac{\partial}{\partial z}\theta - \frac{2e}{c}A_z\right) = 2e|\Psi|^2\ V_{\perp s}. \tag{41}$$

Where, $\boldsymbol{J}_{\parallel s}$ is the maximum supercurrent in the direction parallel to the film surfaces and $J_{\perp s}$ is the maximum supercurrent in the direction perpendicular to the film surfaces. The corresponding electrons supervelocities are

$$\boldsymbol{V}_{\parallel,s} = \frac{1}{m_\parallel}\left(\hbar\boldsymbol{\nabla}_\parallel\theta - \frac{2e}{c}\boldsymbol{A}_\parallel\right), \tag{42}$$

$$V_{\perp,s} = \frac{1}{m_\perp}\left(\hbar\frac{\partial}{\partial z}\theta - \frac{2e}{c}A_z\right) \tag{43}$$

The maximum possible supercurrent that could be carried by a HTS thin film or wire will be attained when the supervelocity $V_i = V_{is}$, (i stands for $\parallel$ or $\perp$)

Minimization of $F_s$ with respect to $|\Psi|^2$ gives the equation

$$M\ f^4 + (1\text{-}M)\ f^2 + \left[\frac{m_\parallel}{2|\alpha|}\left(V_{\parallel,s}^2 + \gamma\,V_{\perp,s}^2\right) - 1\right] = 0\ . \tag{44}$$

The solution of this equation gives

$$f^2 = \frac{1}{2M}\left[ -(1\text{-}M) + \left[ (1\text{-}M)^2 - 4\,M \left[ \frac{m_\parallel}{2|\boldsymbol{\alpha}|}\left( V_{\parallel,s}^2 + \boldsymbol{\gamma}\, V_{\perp,s}^2 \right) - 1 \right]\right]^{1/2} \right] \qquad (45)$$

Note that the positive sign of the square root was chosen for convenience.

Subsequently, the corresponding supercurrents are (cf. eqs. (40) and (41))

$$\boldsymbol{J}_{\parallel s} = \frac{2e}{2M}\, \Psi_e^{\,2}\, \boldsymbol{V}_{\parallel s} \left[ -(1\text{-}M) + \left[ (1\text{-}M)^2 - 4\,M \left[ \frac{m_\parallel}{2|\boldsymbol{\alpha}|}\left( V_{\parallel,s}^2 + \boldsymbol{\gamma}\, V_{\perp,s}^2 \right) - 1 \right]\right]^{1/2} \right] \qquad (46)$$

$$J_{\perp s} = \frac{2e}{2M}\, \Psi_e^{\,2}\, V_{\perp s} \left[ -(1\text{-}M) + \left[ (1\text{-}M)^2 - 4\,M \left[ \frac{m_\parallel}{2|\boldsymbol{\alpha}|}\left( V_{\parallel,s}^2 + \boldsymbol{\gamma}\, V_{\perp,s}^2 \right) - 1 \right]\right]^{1/2} \right] \qquad (47)$$

To obtain maximum $\boldsymbol{J}_{\parallel s}$ and maximum $J_{\perp s}$ we differentiate each of eqs. (46) and (47) with respect to their corresponding velocity $\boldsymbol{V}_{\parallel s}$ and $V_{\perp s}$, one obtain for the velocities

$$V_{\parallel,s} = \left( \frac{|\boldsymbol{\alpha}|}{m_\parallel} \right)^{1/2} \left[ 1 - (1\text{-}M)\, f^2 - M\, f^4 \right]^{1/2} \qquad (48)$$

$$V_{\perp,s} = \left( \frac{|\boldsymbol{\alpha}|}{m_\parallel \boldsymbol{\gamma}} \right)^{1/2} \left[ 1 - (1\text{-}M)\, f^2 - M\, f^4 \right]^{1/2} \qquad (49)$$

Where,

$$\boldsymbol{\gamma}\, V_{\perp,s}^2 = V_{\parallel,s}^2 \quad , \qquad (50)$$

with $\boldsymbol{\gamma} = \frac{m_\perp}{m_\parallel}$.

Substituting eqs. (48) and (49) into eqs. (40) and (41) the maximum supercurrents will have the forms

$$J_{\parallel,s} = 2e\, \psi_e^{\,2} \left( \frac{|\boldsymbol{\alpha}|}{m_\parallel} \right)^{1/2} \left[ 1 - (1\text{-}M)\, f^2 - M\, f^4 \right]^{1/2} f^2 \qquad (51)$$

$$J_{\perp,s} = 2e\, \psi_e^{\,2} \left( \frac{|\boldsymbol{\alpha}|}{m_\parallel \boldsymbol{\gamma}} \right)^{1/2} \left[ 1 - (1\text{-}M)\, f^2 - M\, f^4 \right]^{1/2} f^2 \qquad (52)$$

This will be maximum at that value of $f^2$, which satisfies the following equation,

$$M\, f^4 + \frac{3}{4}\,(1\text{-}M)\, f^2 - \frac{1}{2} = 0$$

From the above we conclude the following result

(A) If $M = 0$, then $f^2 = \frac{2}{3}$ , we get $V_{\parallel,s} = \left(\frac{|\alpha|}{3m_{\parallel}}\right)^{1/2}$, $V_{\perp,s} = \left(\frac{|\alpha|}{3m_{\parallel}\gamma}\right)^{1/2}$, with

$$V_s^2 = V_{\perp,s}^2 + V_{\parallel,s}^2 = \frac{|\alpha|}{3m_{\parallel}}\left(\frac{1}{\gamma} + 1\right) . \qquad (53)$$

For $\gamma = 1$, $V_s^2 = \frac{2|\alpha|}{3m_{\parallel}}$, which corresponds to the isotropic LTS case [18], and the corresponding maximum values of the critical currents read,

$$J_{\parallel,c} = J_{\parallel,s} = 2e\,\psi_e^2\,\frac{2}{3}\left(\frac{|\alpha|}{3m_{\parallel}}\right)^{1/2}$$

$$J_{\perp,c} = J_{\perp,s} = 2e\,\psi_e^2\,\frac{2}{3}\left(\frac{|\alpha|}{3m_{\parallel}\gamma}\right)^{1/2},$$

where ,

$$J_c = \left[J_{\parallel,c}^2 + J_{\perp,c}^2\right]^{1/2} = 2e\,\psi_e^2\,\frac{2}{3}\left(\frac{2|\alpha|}{3m}\right)^{1/2} , \qquad (54)$$

where, $m \equiv m_{\parallel} \equiv m_{\perp}$ ( $\gamma = 1$ )

(B) If $M = \frac{1}{2}$, (the case of HTS), then $f^2 = 0.692$, and we have $V_{\perp,s} = \left(\frac{C\,|\alpha|}{m_{\parallel}\gamma}\right)$ and $V_{\parallel,s} = \left(\frac{C\,|\alpha|}{m_{\parallel}}\right)^{1/2}$, where C = 0.414568.

The square of the supervelocity is

$$V_s^2 = V_{\perp,s}^2 + V_{\parallel,s}^2 = \frac{C\,|\alpha|}{m_{\parallel}}\left(\frac{1}{\gamma} + 1\right)$$

and for the supercurrent components are

$$J_{\parallel,c} \equiv J_{\parallel,s} = 2\,e\,C\,\psi_0^2\left(\frac{|\alpha|}{m_{\parallel}}\right)^{1/2}$$

$$J_{\perp,c} \equiv J_{\perp,s} = 2\,e\,C\,\psi_0^2\left(\frac{|\alpha|}{m_{\parallel}\gamma}\right)^{1/2}$$

$$J_c = \left[ J_{\parallel,c}{}^2 + J_{\perp,c}{}^2 \right]^{1/2} = 2\,e\,C\,\psi_0{}^2 \left( \frac{|\alpha|}{m} \right)^{1/2} \left( \frac{1}{\gamma} + 1 \right)^{1/2}, \qquad (55)$$

where, $J_c$ is the maximum critical current for the HTSs.

### 3.3 Parallel critical field in HTS thin film

To calculate the parallel critical field for HTS thin film with thickness $d < \lambda$, we will consider a sufficiently thin film, in which the applied field may replace the vector potential. Two cases may be considered: the first case concerns the magnetic field parallel to the thin film surface and the second case is that in which the field is perpendicular to the surface. In our coordinate system x is the normal axis to the film surface, which bounded by $x = \pm\ d/2$ planes. Here $H_z \equiv H_\parallel$ and $H_x \equiv H_\perp$; the vector potential $\boldsymbol{A} = (0, Hx, 0)$ for the parallel field and for the $H_\perp$ we take $A_y = |\boldsymbol{A}| = \text{constant } z$.

To find the critical field for a thin film that have second-order phase transition (i.e., $|\psi|^2 \to 0$) and $\lambda_{eff}(H) \to \infty$ then $A_y = Hx$, where $\lambda_{eff}$ in zero field can refers to by $\lambda_\parallel$ with $|\psi|^2 \cong n_s$ - the superelectrons number and H is the external applied magnetic field. Thus $\boldsymbol{v}_s = \frac{-2e}{mc}\boldsymbol{A}$ or $v_{sy} = \frac{-2e}{mc} A_y = \frac{-2e}{mc} H\,x$, and the kinetic energy = $\frac{1}{2}\, m\, v_{sy}{}^2\, |\psi|^2 = \frac{1}{2}\, m \left( \frac{-2\,e\,H\,x}{mc} \right)^2 |\psi|^2$. Therefore the Gibbs free energy per unit area of film is

$$G_s = \int_{-d/2}^{d/2} \left( F_s - \frac{BH}{4\pi} \right) dx \qquad (56)$$

$$G = d\left[ F_n + \alpha(T)|\Psi(r)|^2 + \tfrac{1}{2}\beta(T)|\Psi(r)|^4 + \tfrac{1}{3}g_o|\Psi(r)|^6 - \frac{H^2}{8\pi} \right] + \\ + \frac{e^2 d^3 H^2}{6m_\parallel^* c^2}|\Psi(r)|^2 + \int_{-d/2}^{d/2} \frac{(B-H)^2}{8\pi} dx \qquad (57)$$

As it was mentioned above, in thin film we can approximate B by H and therefore the last term in eq. (57) may be dropped. Thus,

$$G = d\left[ F_n + \alpha(T)|\Psi(r)|^2 + \tfrac{1}{2}\beta(T)|\Psi(r)|^4 + \tfrac{1}{3}g_o|\Psi(r)|^6 - \frac{H^2}{8\pi} \right] + \frac{e^2 d^3 H^2}{6m_\parallel^* c^2}|\Psi(r)|^2 \qquad (58)$$

Minimizing G with respect to $|\psi|^2$, we get

$$\alpha(T)+\beta(T)\left|\Psi(r)\right|^2+g_o\left|\Psi(r)\right|^4+\frac{e^2d^2H^2}{6m_{\parallel}^*c^2}=0\,. \qquad (59)$$

Using the results of sec.2 it is easy to find

$$\left|\Psi\right|^2=\Psi_e{}^2\,\frac{-(1\text{-}M)\pm\sqrt{(1\text{-}M)^2-4M\left(\dfrac{e^2d^2H^2}{6m_{\parallel}c^2|\alpha|}-1\right)}}{2M} \qquad (60)$$

In the normal phase $|\psi|^2\rightarrow 0$, and this takes place if M=1 and $\frac{e^2d^2H^2}{6m_{\parallel}c^2|\alpha|}=1$. By using that[21]

$$\alpha=-\frac{2e^2H_c{}^2\lambda_{\parallel}^2(T)}{\left(1+\dfrac{M}{3}\right)m_{\parallel}c^2}\quad , \qquad (61)$$

we get,

$$H_{\parallel}=\frac{2\sqrt{6}\;H_c\lambda_{\parallel}(T)}{d\left(1+\dfrac{M}{3}\right)^{1/2}} \qquad (62)$$

The last formula indicates that the parallel critical field may exceed the thermodynamic critical field $H_c$ . The physical reason for this is simply that the thin film, being largely penetrated by the field, has little diamagnetic energy for given applied field in comparison to an equal volume of a bulk superconductor [18].

## Discussions and Results

1. The MGL equations were used to solve three fundamental problems for HTS. The first one was the domain wall energy. It was shown that the surface energy, in this case, has negative sign, which emphasize that this partition is energetically favored in spite of the existence of the sixth power term of the order parameter. Our results show that the modification in the GL free energy function is important to meet with the new characteristics of the novel materials.

2. The second problem (sec. (3.2)) concerns the calculation of the maximum supercurent in HTS thin film. It was shown that the supercurrent in the direction normal to the surface of the SC sample depends on the anisotropy factor $\gamma$. If the anisotropy factor is very large the normal current component will tend to zero (see eq. (54.1)). One the other hand, the parallel component does not depend on $\gamma$, and therefore, the net value of the super current, eq. (55), for large values of $\gamma$ will have large value compared with that derived from the usual GL equations (compare with ref.[19]). The maximum supercurrent also depends on the magnetic field and on the coherence length, the phase composition of the sample, e.g., single or multiphase, and on the orientation of the external field with respect to the layered planes.

3. The parallel to the SC surface critical magnetic field in HTS thin film is calculated in sec. (3.3). It was shown that the parallel critical magnetic field that was calculated by the MGL equations depends on the parameter M, which characterizes the sixth power of the order parameter in the MGL free energy functional.

4. In our work less concern was given to the role of the surface boundary conditions in the MGL free energy variation procedure. This term was put to be zero, as was usually doing in driving GL equations. However, such a problem needs careful study as it is shown in ref.[22]. It was shown that the free energy variation depends on whether the variation is going with respect to the vector potential $\boldsymbol{A}$ or with respect to the magnetic field itself.